**Beyond Physicians: Social and Familial Norms Driving Cesarean Section Decisions in Bangladesh**

Jamal Uddin

*Department of Communication, Cornell University, Ithaca, USA*

**Author Note**

Correspondence concerning this article should be addressed to Jamal Uddin, Dept. of Communication, 60 Mann Library, 237 Mann Drive, Ithaca, NY-14850, USA. Email: jamal.uddin@cornell.edu

**Abstract**

Women's health in Bangladesh faces risks due to an alarming rise in cesarean section (CS) rates, exceeding 72% in hospital-based deliveries—far surpassing the WHO's recommended limit of 15%. This study, guided by the Health Belief Model (HBM) and the Theory of Planned Behavior (TPB), explored socio-cultural factors influencing childbirth mode decisions. Among 503 survey participants, 91% of CS cases occurred against initial preferences, revealing a disconnect between health beliefs and behavior. Subjective norms, particularly family influence and social expectations, emerged as more critical in shaping CS decisions than physician recommendations.

**Beyond Physicians: Social and Familial Norms Driving Cesarean Section Decisions in**

Unnecessary CS poses a significant health threat to expectant women in Bangladesh, as well as globally. Over the last 15 years, unnecessary CS in Bangladesh has increased eightfold, primarily due to the rise in hospital-based deliveries (Doraiswamy et al., 2020). Various studies reflect this growing trend, with CS rates as high as 82.9% in mostly private facilities (Alam et al., 2017; Begum et al., 2018). The medically indicated CS rate, however, remains low. For instance, Begum (2018) found that only 1.4% of CS deliveries were medically necessary. In Bangladesh, CS is more prevalent among women with higher socio-economic status (SES) and better antenatal care (Begum, 2018). This raises a critical question: Why do women of higher SES opt for unnecessary CS, despite the known risks to both mothers and babies?

While previous research has examined CS prevalence, its correlation with socioeconomic factors (Karim et al., 2020), physician-patient communication (Doraiswamy et al., 2020), and its economic burden (Haider et al., 2018), few studies have focused on the health beliefs and socio-cultural factors that influence women's childbirth decisions. This study aims to address that gap by applying two widely used health behavior theories—the Health Belief Model (HBM) and the Theory of Planned Behavior (TPB).

Research on preventive health shows that individuals' health perceptions strongly influence their decisions to take action. In this context, risk perception—central to HBM—plays a crucial role in shaping health behaviors (Rosenstock, 1974). Meanwhile, TPB's constructs of attitude, subjective norms, and perceived behavioral control are key predictors of an individual's intention to engage in specific behaviors (Ajzen, 1991). Together, these theories provide a comprehensive framework for understanding Bangladeshi women's health perceptions, the social factors shaping their decisions, and their behavior regarding CS.

*Decision-making about birth delivery mode*

Women's birth delivery decisions are influenced by various factors, including demographic characteristics, socioeconomic status, knowledge, expectations, birth complications, and concerns for maternal and infant health (Coates et al., 2020; Goodall et al., 2009; Loke et al., 2015). Negative birth experiences and poor patient-provider communication also play a role (Aktaş & Aydın, 2019). Common non-medical reasons for CS across countries include safety perceptions, fear of pain, past experiences, healthcare guidance, social norms, and access to information (Coates et al., 2020; Kumar & Lakhtakia, 2021). Both personal beliefs and social influences significantly impact birth delivery choices in high- and low-income settings.

Loke et al. (2015) found that higher education levels led women in Hong Kong to favor vaginal delivery (VD) (62.2%) over unnecessary CS (43.8%). However, studies from Iran and Bangladesh showed that women with higher education and incomes were more likely to prefer CS (Darsareh et al., 2016; Doraiswamy et al., 2020). CS rates were higher in private hospitals, where wealthier individuals tend to seek care. In Turkey, Buyukbayrak et al. (2010) found that factors like early recovery and prior VD history favored VD, while fear of labor pain and requests for tubal ligation were linked to CS. However, education, occupation, and gestational age did not significantly influence delivery mode, suggesting that preferences vary by country, with elective CS more common in developing nations (Buyukbayrak et al., 2010).

*Health belief model (HBM)*

The HBM suggests that perceived susceptibility, severity, and efficacy drive health behavior changes (Rosenstock, 1974). Without awareness of health risks, individuals are unlikely to engage in preventive behavior, making effective communication essential (Rosenstock, 1974; Becker et al., 1977). Key constructs of the model include perceived susceptibility, severity, benefits, barriers, cues to action, and self-efficacy. Perceived susceptibility is the belief in one's vulnerability to a health condition, while perceived severity reflects the seriousness of that condition (Janz & Becker, 1984). Perceived benefits represent the advantages of preventive action, and perceived barriers include obstacles like cost or inconvenience. Cues to action motivate individuals to take preventive measures, and self-efficacy is the confidence in one's ability to act.

Darsareh et al. (2016) found that Iranian women with lower self-efficacy preferred CS, while those with higher self-efficacy chose VD. However, perceived benefits and barriers showed no significant differences in birth modality decisions. In Bangladesh, reliance on physicians and financial incentives may lead to higher CS rates, without fully discussing the risks (Darsareh et al., 2016). While HBM provides insights into health behavior, not all constructs are equally effective in predicting actions. Carpenter (2010) found that perceived benefits and barriers are the strongest predictors, especially in preventive health. However, external social and environmental factors also influence health decisions (Green et al., 2020). Combining theories like HBM and TPB offers a more comprehensive understanding of behavior (Yang, 2015).

### *Theory of planned behavior (TPB)*

The TPB explains health behavior through attitude, subjective norms, and perceived behavioral control (Ajzen, 1991). Intentions are shaped by attitudes (positive or negative evaluations), social pressures (subjective norms), and perceived control over behavior. TPB has been widely used in health contexts, such as smoking cessation (Norman et al., 1999), young adults' H1N1 vaccine intentions (Yang, 2015), and chronic illness treatment adherence (Rich et al., 2015). Norman et al. (1999) found TPB explained nearly 50% of the variance in smokers' intentions to quit, with perceived control as the strongest predictor. In chronic illness studies, attitude, subjective norms, and perceived control explained 33% of adherence intentions (Rich et al., 2015). The theory's application across health contexts shows its effectiveness in understanding behavioral intentions, though the strength of constructs can vary. Adding factors like perceived susceptibility can enhance TPB's predictive power (Norman et al., 1999).

*Doctors' influence as a subjective norm*

Doctors' guidance and fear of labor pain influence women's birth decisions (Betrán et al., 2018). Research shows that support from healthcare professionals and family members significantly affects women's delivery preferences (Shams-Ghahfarokhi & Khalajabadi-Farahani, 2016). Physicians' communication, particularly regarding labor pain or emergency interventions, can create uncertainty in decision-making (Goodall & Mcvittie, 2009). In Bangladesh, physician communication plays a crucial role in the rising CS rates (Doraiswamy et al., 2020). Thus, examining physicians' influence as a distinct aspect of subjective norms is essential for understanding their impact on women's birth delivery decisions.

Based on the above theoretical discussion, this study is structured around the framework below (Figure 1), guided by a research question and a total of seven hypotheses.

**Figure 1.** An integrated framework of the study.

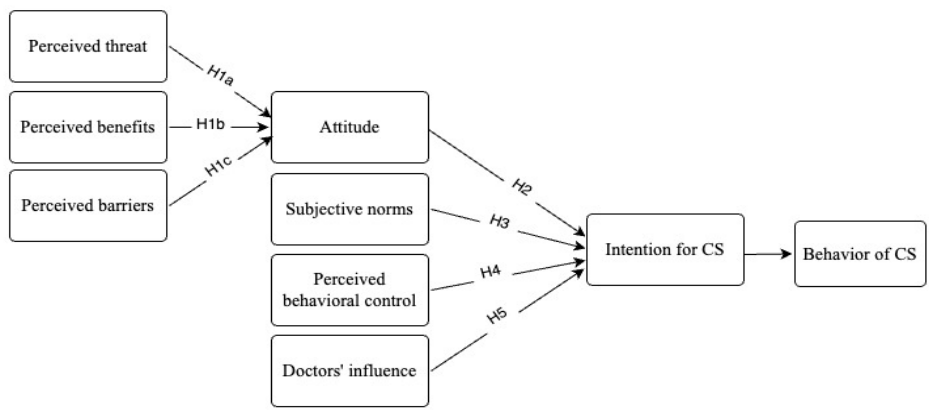

*Research question*

RQ1: Which major factors influence expecting mothers in Bangladesh to decide their birth mode?

*Hypotheses*

**H1**: Components of the Health Belief Model (HBM) (perceived health threat, benefits, and barriers) influence women's attitudes toward birth delivery intentions.

> **H1a**: Greater perceived health threat from CS leads to a more negative attitude toward CS.
>
> **H1b**: Greater perceived health benefits of CS lead to a more positive attitude toward CS.
>
> **H1c**: Greater perceived health barriers to CS lead to a more negative attitude toward CS.
>
> **H2**: A more positive attitude toward CS is associated with a higher intention to deliver via CS.
>
> **H3**: Stronger subjective norms favoring CS are associated with a higher intention to deliver via CS.
>
> **H4**: Greater perceived behavioral control over CS is associated with a higher intention to deliver via CS.
>
> **H5**: Physicians positively influence women's birth delivery intentions toward CS.

**Method**

*Study design*

This study used an online survey via Qualtrics to collect data from Bangladeshi women aged 18 to 35 who had given birth within the past two years. A combination of convenience sampling (e.g., recruitment through pregnancy-related Facebook groups) and snowball sampling was employed to identify and recruit participants who provided consent and were available. The survey explored participants' preferred birth delivery mode and the factors influencing their choices, focusing on the four constructs of the HBM and the three constructs of the TPB. This anonymous cross-sectional study was approved by [BLINDED] and conducted between May 23 and June 5, 2022.

The survey items were translated into Bangla to ensure inclusivity and accessibility. A pilot test with 15 participants assessed readability, response time, and cultural relevance. To boost participation, every third respondent was randomly awarded a small incentive of BDT 200 (approximately $2.30).

*Questionnaire development*

The survey was designed based on previous studies that applied TPB and HBM frameworks or focused on birth delivery mode preferences. It consisted of three sections. The first section gathered information on actual behavior, birth intention (without medical necessity), attitudes, perceived behavioral control, and subjective norms (Ajzen, 2019; Buyukbayrak et al., 2010; Glanz et al., 2008; Loke et al., 2015; Shams-Ghahfarokhi & Khalajabadi-Farahani, 2016). The second section collected data related to the HBM constructs—perceived susceptibility, severity, benefits, and barriers—regarding VD and CS (Loke et al., 2015). The final section collected demographic information from the participants.

*Measures*

*Actual Behavior and Birth Intention*

Actual birth modality was assessed with a binary question: "I delivered my baby via vaginal delivery (VD) or cesarean section (CS)." Birth modality intention prior to delivery was measured using the statement, **"Before delivery, my plan was CS or VD,"** rated on a Likert scale where 1 = did not intend to, 2 = somewhat intended to, and 3 = very much intended to.

*Attitude*

Attitudes toward CS were measured using 10 items designed to capture participants' positive or negative perceptions. Statements such as "Cesarean is more comfortable than vaginal delivery" were rated on a five-point Likert scale (strongly agree to strongly disagree), with total scores ranging from 10 to 50. Higher scores indicated a more favorable attitude toward CS.

*Subjective norms*

Subjective norms were assessed through 6 items that examined the influence of people and media on birth decisions and perceived social norms around CS. Statements like "Nowadays, doctors recommend a cesarean section" were rated, with total scores ranging from 6 to 30. Higher scores reflected greater perceived social pressure toward CS.

*Doctors' influence as a subjective norm*

Doctors' influence was measured using four items, including statements like, "Although I planned for a vaginal delivery, my doctor asked me to prepare for a C-section to avoid any risks with VD." Higher scores indicated stronger doctor influence in birth decision-making.

*Perceived behavioral control*

Perceived behavioral control was measured using three items: "I was confident I could decide on my delivery mode," "Deciding about my delivery mode was easy for me," and "Deciding about the delivery mode was not entirely up to me" (reverse-coded).

*HBM constructs*

The HBM constructs—perceived susceptibility, severity, benefits, and barriers—were assessed using the prompt, "How much do you agree with the following statement?" Perceived susceptibility and severity were combined as perceived threat to better capture health beliefs about birth mode. Each construct was measured with 10 items for both VD and CS, rated on a five-point scale (1 = strongly disagree to 5 = strongly agree), with total scores ranging from 10 to 50. Higher scores reflected stronger perceptions of each construct.

*Demographic variables*

Demographic information collected included participants' age, education, employment status, household income, number of antenatal care visits, time since last childbirth, delivery location (hospital or home), living location (urban or rural), and racial identity.

*Data analysis*

A descriptive statistical analysis was conducted in two parts: the first focused on demographic variables, and the second analyzed theoretical constructs from the HBM and TPB.

Structural equation modeling (SEM) was then used to examine participants' intentions regarding CS. SEM integrates path analysis and confirmatory factor analysis (CFA) to explore relationships between variables (Fan et al., 2016). Path analysis evaluates direct and indirect relationships, while CFA identifies latent constructs by analyzing shared variance among related variables (Fan et al., 2016; Suhr, 2006).

The SEM process involved five steps: model specification, identification, parameter estimation, model evaluation, and modification (Fan et al., 2016; Suhr, 2006). SEM was performed using the R programming language with the Lavaan package.

*Sample size*

There are two primary methods to determine the sample size for SEM: Monte Carlo data simulation and the rule of thumb. Monte Carlo simulations recommend sample sizes ranging from 30 to 460, depending on factors such as the number of indicators, factor loadings, and factors (Wolf et al., 2013). The rule of thumb suggests 200 to 400 participants, typically 10 to 20 times the number of variables.

This study, with 503 participants, exceeds both Monte Carlo and rule-of-thumb recommendations, meeting the criteria for an effect size of 0.5, power of 0.8, and a p-value of 0.05. However, SEM sample size requirements also depend on factors like fit indices, model size, variable distribution, missing data, reliability, and path coefficients (Fan et al., 2016; Wolf et al., 2013).

**Results**

A total of 702 participants attempted the online survey, but only 598 met the eligibility criteria. Of these, 509 completed the entire survey. After identifying and removing outliers based on age (over 35), the final sample size was reduced to 503. The results section is divided into three parts: descriptive analysis of participants' demographic characteristics, answers to the research question, and hypothesis testing of the conceptual model.

The demographic data include seven key variables: age, education, income, employment, doctor visits during pregnancy, birthplace, and living location (see Table 1). Of the 503 participants, 366 (73%) delivered via cesarean section (CS), while 137 (27%) had a vaginal delivery (VD). Notably, 85% of deliveries in private hospitals were CS, compared to 46% in public hospitals.

Regarding education, 32% of participants had a master's degree or higher, 38% held a bachelor's degree, and 25% had completed 12th grade or a diploma. Among those who had a CS, 34% had a master's degree or higher, 39% had a bachelor's degree, and 22% had completed 12th grade or a diploma. In contrast, for those who had a VD, 26% had a master's degree or higher, 35% held a bachelor's degree, and 31% had completed 12th grade or a diploma. These findings suggest that CS was more common among women with higher levels of education. A chi-square test revealed a significant association between education and birth delivery mode ($x^2(df=4) = 9.25, p \geq .05$).

**Table 1.** Demographic variables.

| Variables | Categories | Cesarean delivery(%) n=366 | Vaginal delivery(%) n=137 | Total % |
|---|---|---|---|---|
| Age | | | | |
| | 18-20 | 22 (0.06) | 10(0.07) | 6 |
| | 21-23 | 73 (0.20) | 34(0.25) | 21 |
| | 24-26 | 87(0.24) | 37(0.27) | 25 |
| | 27-29 | 92(0.25) | 31(0.23) | 24 |
| | 30-32 | 65(0.18) | 14(0.10) | 16 |
| | 33-35 | 20(0.05) | 11(0.08) | 6 |
| Education | | | | |
| | Primary (Till 5th grade) | 2 | 1 | 1 |
| | High school (Till 10th grade) | 19 (0.05) | 10(0.07) | 6 |
| | College (Till 12th grade) | 47(0.13) | 31(0.23) | 16 |
| | Vocational education (Diploma) | 32(0.09) | 11(0.08) | 9 |
| | Above college (bachelor's degree) | 141(0.39) | 48(0.35) | 38 |
| | Higher education (Masters, MPhil, Ph.D.) | 125(0.34) | 36(0.26) | 32 |
| Income | | | | |
| | Tk 0-Tk15,000 ($166) | 74(0.20) | 49(0.36) | 26 |
| | Tk15,001-Tk30,000 ($332) | 100(0.27) | 19(0.14) | 25 |
| | Tk30,001-Tk45,000 ($498) | 76(0.21) | 22(0.16) | 20 |
| | Tk45,001-Tk60,000 ($666) | 43(0.12) | 12(0.09) | 11 |
| | Tk60,001-Tk75,000 ($833) | 26(0.07) | 4(0.03) | 6 |
| | Tk75,0001 and above | 37(0.10) | 19(0.14) | 12 |
| Employment | | | | |
| | Fulltime | 90(0.25) | 18(0.13) | 22 |
| | Parttime | 19(0.05) | 5(0.04) | 5 |
| | Unemployed | 18(0.05) | 6(0.04) | 5 |
| | Housewife (Unemployed also) | 239(0.65) | 108(0.79) | 69 |
| Doctor visits | | | | |
| | 1 visit | 2 | 2 | 1 |
| | 2 visits | 9(0.02) | 9(0.07) | 4 |
| | 3 visits | 18(0.05) | 22(0.16) | 8 |
| | 4 visits | 47(0.13) | 16(0.12) | 13 |
| | 5 visits | 40(0.11) | 17(0.12) | 11 |
| | 6 visits | 53(0.14) | 13(0.09) | 13 |
| | 7 or 7+ visits | 197(0.54) | 58(0.42) | 51 |
| Birthplace | | | | |
|  | Public hospital | 36(0.10) | 43(0.31) | 15 |
|  | Private hospital | 320(0.87) | 55(0.40) | 75 |
|  | Health center | 8(0.02) | 16(0.12) | 5 |
|  | Home | 1 | 23(0.17) | 5 |
| Living location |  |  |  |  |
|  | Capital | 131(0.36) | 37(0.27) | 34 |
|  | Divisional City | 37(0.10) | 15(0.11) | 10 |
|  | District Level City | 85(0.23) | 27(0.20) | 22 |
|  | Thana Level City | 55(0.15) | 17(0.12) | 14 |
|  | Village | 58(0.16) | 40(0.29) | 20 |

*CS= Cesarean section, *VD= Vaginal delivery, *n=total number, $1=BDT90

*Major factors influencing birth delivery decisions*

To address the research question on the influence of birth delivery mode, the theoretical constructs of the TPB and HBM were analyzed to examine their impact on birth modality intentions.

The TPB consists of four constructs, each applied in this study to explore their influence on Bangladeshi mothers' birth decisions. Results for the intention variable indicate that only 9% of participants strongly favored CS, while 8% of mothers who had a vaginal delivery (VD) expressed dissatisfaction with it (see Table 2). The mean attitude score for VD (e.g., "vaginal delivery is safer for both mother and baby") was 25 out of 30, compared to 26 out of 50 for CS (e.g., "cesarean delivery is more comfortable than vaginal delivery"), reflecting lower intention toward CS.

Subjective norms had a greater influence on VD (mean score 13.53 out of 20) than on CS (mean score 13.41 out of 30). However, the influence of doctors, measured separately under subjective norms, was strong for both delivery modes, with slightly higher influence for CS (mean score 14.57 out of 20) compared to VD (mean score 13.96 out of 20).

Perceived behavioral control (PBC) was higher among mothers who had a VD (mean score 11.08 out of 15) compared to those who had a CS (mean score 9.43 out of 15). An independent t-test confirmed a significant difference in PBC between the two groups ($t_{(df=501)}=7.94; p<.001$).

**Table 2.** Health behavior intention.

| Actual behavior | Did not intend to n(%) | Somewhat intend to n(%) | Very much intend to n(%) |
|---|---|---|---|
| VD | 11 (8) | 13 (10) | 113 (82) |
| CS | 224 (61) | 110 (30) | 32 (9) |

All three HBM constructs—perceived health threats (comprising perceived susceptibility and severity), health benefits, and health barriers—were examined to understand participants' health beliefs regarding CS and VD. A dependent t-test revealed a significant difference in health beliefs based on birth type (see Table 3).

**Table 3.** Health beliefs about the birth consequences.

| Level of health beliefs | M | SD | *t* | Sig. |
|---|---|---|---|---|
| Health threats (CS) | 15.37 | 3.22 | -13.25 | .001 |
| Health threats (VD) | 13.47 | | | |
| Health benefits (CS) | 11.02 | 2.80 | 16.28 | .001 |
| Health benefits (VD) | 13.05 | | | |
| Health barriers (CS) | 10.98 | 2.74 | 2.67 | .004 |
| Health barriers (VD) | 11.31 | | | |

*Structural equation modeling (SEM)*

The SEM models were employed to test the proposed hypotheses regarding the preference for CS. After data cleaning, 503 participants were included in the analysis, of which 366 had undergone CS. The 137 participants who had VD were excluded. With a sample size of 366, the study met the requirement for reliable SEM estimates, as a minimum of 200 participants is considered adequate (Kline, 2015).

The SEM model development followed several stages: data cleaning, testing for normality, checking for common method bias, building the measurement model with exploratory factor analysis (EFA), conducting CFA, assessing reliability, and estimating the SEM. Analysis was conducted using the Lavaan package in the R programming environment.

Of the dataset's 106 variables, 34 relevant variables were selected for the model. To improve model fit, 20 items with factor loadings below 0.50 were removed, retaining only significant items ($p < .001$). Reliability was confirmed, with all measurement items achieving Cronbach's Alpha values above the recommended threshold (Hair et al., 2010). The final measurement items are presented in the Table 4.

**Table 4.** Measurement items and their reliability for CS model

| Constructs and their respective items | Cronbach's α | Factor loadings |
|---|---|---|
| **Intention related to CS (INTC)** | | |
| From the beginning, my birth modality intention was CS | 0.76 | 0.655 |
| **Doctor influence for CS (DIC)** | | |
| -For the sake of my baby's well-being, the doctor or midwife strongly recommends me to have a cesarean section. | 0.76 | 0.857 |
| - For my health, the doctor or midwife strongly recommends me to have a cesarean section. | 0.77 | 0.902 |
| - After consulting with my doctor or midwife, my decision for birth modality changed from vaginal delivery to cesarean section. | 0.77 | 0.724 |
| - *Doctors have had no role in my decision about the choice of delivery* | 0.77 | *dropped* |
| **Subjective norms related to CS (SNC)** | | |
| - Today, all women do a cesarean section. | 0.75 | 0.688 |
| - *Nowadays, doctors recommend a cesarean section.* | *0.77* | *dropped* |
| - At a time when there are many technologies, why I should bear the pain of vaginal childbirth. | 0.75 | 0.725 |
| - I will do a cesarean section because we have had a cesarean experience in our family, without any problem. | 0.74 | 0.826 |
| - If I do a vaginal delivery, I will be embarrassed in front of my relatives. | 0.76 | *dropped* |
| - Those who are not well off would do vaginal delivery. | 0.76 | *dropped* |
| **Attitude related to CS (ATTC)** | | |
| -Cesarean is more comfortable than vaginal delivery. | 0.75 | 0.623 |
| -Because I am financially solvent, I am not worried about the cost of a cesarean section. | 0.75 | 0.581 |
| - *Because of anesthesia during a cesarean, I am more comfortable.* | 0.75 | *dropped* |
| - Children who are born via cesarean section are smarter. | 0.76 | 0.535 |
| -Preserve sexual function and genital appearance | 0.75 | *0.561* |
| - *Cesarean delivery prevents neonatal death.* | 0.76 | *dropped* |
| - *Due to the inappropriate behavior of hospital staff during the hours of labor, I prefer cesarean.* | 0.76 | *dropped* |
| - Medical care during cesarean is better than during vaginal delivery. | 0.76 | 0.591 |
| - Those who want only one or two children are better to do cesarean. | 0.76 | *dropped* |
| - *Previous cesarean section history* | 0.75 | |
| | 0.76 | |
| **Perceived behavioral control (PBC)** | | |
| -I was confident that I could decide about my delivery mode | 0.76 | *0.599* |
| -For me to decide about my delivery mode was easy | 0.76 | 0.727 |
| -*Deciding delivery mode was not entirely dependent on me* | | *dropped* |
| | 0.77 | |
| **Health threat related to CS (HTC)** | | |
| -Abdominal wound infection | 0.77 | *dropped* |
| -Long recovery time | 0.76 | 0.548 |
| -Concern over the anesthesia complications of CS | | 0.593 |
| *-Afraid of uterine scar ruptures if a cesarean delivery is performed* | 0.77 | *dropped* |
| | *0.77* | |
| **Health benefits related to CS (HBC)** | | |
| -Less fear of prolonged labor and fetal injuries | 0.76 | 0.508 |
| -Avoid pain induced by repetitive vaginal examinations | 0.77 | 0.516 |
| -Avoids the uncertainty of the timing of the delivery | | 0.526 |
| | 0.76 | |
| **Health barriers related to CS (HBrC)** | | |
| -Cost of CS | 0.77 | 0.718 |
| *-Cannot choose CS in a public hospital* | 0.77 | *dropped* |
| *-Religious/traditional belief* | | *dropped* |
| | 0.77 | |

CFA model-fit: χ2 (169) = 326.347, CFI = 0.90, TLI = 0.87, RMSEA = 0.05, SRMR = 0.07; Standard Alpha represents the value of Cronbach's alpha.

The measurement model demonstrated a good fit, with Comparative Fit Index (CFI) scores above the recommended 0.90 threshold and Root Mean Square Error of Approximation (RMSEA) and Standardized Root Mean Square Residual (SRMR) scores below the recommended 0.08 (Hair et al., 2010). However, the Tucker-Lewis Index (TLI) was slightly below the minimum threshold at 0.88. After completing all steps, the study's measurement model was successfully established (see Table 5).

**Table 5.** Descriptive statistics and correlations among latent variables of CS model.

| Latent variables | *M* | *SD* | INT | DIC | SNC | AttC | PBC | HTC | HBC | HBrC |
|---|---|---|---|---|---|---|---|---|---|---|
| **INTC** | **1.48** | **.65** | 1 | | | | | | | |
| **DIC** | **3.91** | **1.12** | 0.109 | 1 | | | | | | |
| **SNC** | **2.04** | **0.92** | 0.548 | 0.095 | 1 | | | | | |
| **AttC** | **2.30** | **1.04** | 0.294 | 0.19 | 0.653 | 1 | | | | |
| **PBC** | **2.98** | **1.08** | 0.188 | 0.127 | 0.21 | 0.359 | 1 | | | |
| **HTC** | **4.1** | **.77** | -0.13 | -0.013 | -0.15 | -0.244 | -0.073 | 1 | | |
| **HBC** | **3.66** | **.87** | 0.16 | 0.093 | 0.25 | 0.275 | 0.171 | -0.031 | 1 | |
| **HBrC** | **4.26** | **.72** | -0.095 | 0.018 | -0.162 | -0.194 | -0.133 | 0.345 | 0.045 | 1 |

With the measurement model established, the study proceeded to structural modeling to examine associations between latent variables. Due to non-normal data, maximum likelihood estimation with robust standard errors (MLR) was used, as recommended by Rosseel (2012). Complex SEM models with over 12 items often face challenges in matching theoretical and observed structures (Hair et al., 2010). A good model fit is determined by a chi-square to degrees of freedom (DF) ratio below three, which this study achieved with a ratio of 1.45 for the CS group, indicating a good fit (Bollen & Long, 1992).

Figure 2 presents the SEM model, and Table 6 summarizes the hypothesis testing results for the CS model. Two hypotheses were found to be significant: (H1b) and (H3). Additionally, two other hypotheses were weakly supported: (H1a) and (H1c).

**Figure 2.** Path diagram of latent variables for CS model.

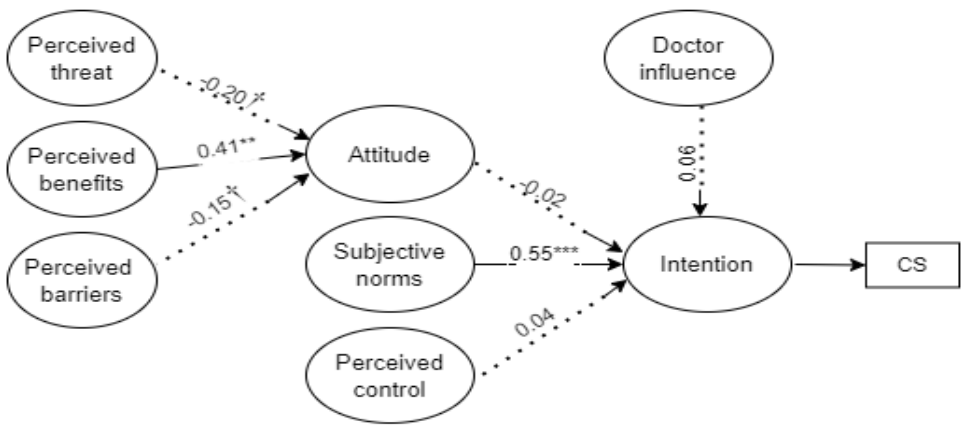

**Table 6.** Summary of hypothesis testing for CS model.

| Hypothesis | Standardized coefficient | Remark |
|---|---|---|
| H1a: Greater the health threat → negative attitude toward CS | -0.201(0.060)† | Weakly supported |
| H1b: Greater perceived health benefits → positive attitude toward CS | 0.408(0.003)** | Supported |
| H1c: Greater the health barrier → negative attitude toward CS | -0.154(0.090) † | Weakly supported |
| H2: Positive attitude towards CS → higher intention for CS | -0.019(0.812) | Rejected |
| H3: A more positive influence of SN toward CS → higher intention for CS | 0.547(0.000)*** | Supported |
| H4: A high perceived behavioral control toward CS → higher intention for CS | 0.044 (0.746) | Not supported |
| H5a: Doctor positive influence for CS → higher intention for CS | 0.057(0.360) | Not supported |

†p < 0.10, **p < 0.01, ***p < 0.001.

**Discussion**

The alarming rise in cesarean section (CS) rates in Bangladesh, which poses long-term health risks to women, remains understudied regarding health beliefs and social factors influencing birth decisions. This study, guided by the HBM and TPB frameworks, explores women's health behavior related to childbirth to address this gap.

CS rates were significantly higher among women with higher income (78%), education (75%), and frequent doctor visits (70%), consistent with trends in other Asian countries (Doraiswamy et al., 2020; Rai et al., 2017; Verma et al., 2020). Private hospitals in Bangladesh exploit this trend, with CS being 13 times more frequent in private rural hospitals than in public facilities, despite only 1.4% of procedures being medically necessary (Begum, 2018). Family pressure on physicians further exacerbates the situation (Dey, 2017; Doraiswamy et al., 2020).

Women in higher SES groups in Bangladesh are more likely to undergo CS, contrasting with similar groups in developed countries, where VD is preferred. This discrepancy may stem from women's lower social positions, dependence on family members and physicians, and the social prestige linked to CS and private hospitals. The findings indicate that doctors, private hospitals, and societal norms are key contributors to Bangladesh's rising CS rates.

The study revealed that 61% of women who underwent CS did not initially desire it, reflected in their low attitude scores (26 out of 50). Health belief scores also indicated a preference for VD, with higher perceived threats (30.84 vs. 27.03) and perceived benefits (22.21 vs. 26.02) for VD. Doctor influence under subjective norms (14.57 vs. 13.96) played a significant role in CS decisions.

Women's birth choices are shaped by multiple factors. While most participants preferred VD, the actual CS rate remained high, with doctor influence being a stronger factor than other social influences. Financial incentives for doctors and hospitals, coupled with poor antenatal care quality, likely contribute to this trend (Doraiswamy et al., 2021).

SEM analysis found significant support for health benefits influencing attitudes toward CS (H1b) and subjective norms driving CS intention (H3). Health threats (H1a) and barriers (H1c) were weakly supported, while perceived behavioral control (H4) and doctor influence (H5) had limited impact. Most participants viewed CS negatively, but 91% of those who underwent CS did so despite their preferences, highlighting a disconnect between health beliefs and behavior.

In Bangladesh, women often lack autonomy, relying on family members for childbirth decisions (Shabuddin et al., 2017). Families depend on physicians, who may prioritize financial incentives over patient welfare, often recommending CS without fully explaining risks. Subjective norms, defined as social pressure from others, significantly influenced CS decisions, explaining 55% of the variance.

These findings align with studies in Iran, where rising CS rates are driven by social trends, doctor recommendations, and perceptions that VD is for women with lower financial means (Naghibi et al., 2021; Shams-Ghahfarokhi & Khalajabadi-Farahani, 2016). However, this study found only modest support for doctor influence (H5) and no significant role for perceived behavioral control (H4), likely due to women's dependency on family members and limited access to accurate information. If women were given to information, emotional, and decision making support, there is a high possibility of increasing their health behavioral control (Farnworth et al., 2008).

The concept of safe delivery remains poorly understood among Bangladeshi women, regardless of SES or education. Promoting healthcare professionalism and raising awareness among families—especially mothers and husbands—is crucial to reducing unnecessary CS rates. While CS is essential in certain medical situations, overuse can lead to long-term complications, higher costs, and reduced mother-baby bonding.

Culturally tailored health campaigns and physician training, similar to successful initiatives in India and China (Singh et al., 2018; BBC, 2019), could promote maternal health and encourage safe delivery practices. Subjective norms must account for family and cultural values in Bangladesh to be effective.

The findings suggest that TPB may be less effective in explaining health behavior in collectivist cultures like Bangladesh, where family members and doctors significantly influence decisions (Doraiswamy et al., 2020; Shabuddin et al., 2017). Models that incorporate cultural dynamics, such as Dutta's (2008) approach, may provide deeper insights into health behavior.

This study's reliance on online recruitment, primarily via Facebook, may not represent the broader population. Additionally, data from women who gave birth within the last two years may be affected by recall bias. Future research should explore offline populations and examine doctors' perspectives to address communication gaps about CS risks.

**Conclusion**

This study theoretically demonstrated that subjective norms significantly influence CS delivery. Not only do physicians have a direct impact, but family members, such as pregnant women's mothers and husbands, who traditionally hold high social positions, also play a significant role in

**Cesarean Trends in Bangladesh**                                                                 21shaping pregnant women's birth decisions. This study categorized and defined social influence in terms of cultural values and subjective norms, highlighting the importance of focusing on cultural values to promote health behavior change among the pregnant women.

**References**

Ajzen, I. (2019). Theory of planned behavior diagram.

   https://people.umass.edu/aizen/tpb.diag.html

Ajzen, I. (1991). The theory of planned behavior. *Organizational Behavior and Human Decision Processes, 50*(2), 179-211. https://doi.org/10.1016/0749-5978(91)90020-T

Aktaş, S., & Aydın, R. (2019). The analysis of negative birth experiences of mothers: A qualitative study. *Journal of Reproductive and Infant Psychology*, *37*(2), 176-192. https://doi.org/10.1080/02646838.2018.1540863

Alam, S., Haider, M., Alam, N., & Nahar, Q. (2017, November). Elevation of cesarean delivery in Bangladesh: a ten-year follow up study. In *2017 International Population Conference*. IUSSP.

Bauserman, M., Lokangaka, A., Thorsten, V., Tshefu, A., Goudar, S. S., Esamai, F., ... & Bose, C. L. (2015). Risk factors for maternal death and trends in maternal mortality in low-and middle-income countries: a prospective longitudinal cohort analysis. *Reproductive Health, 12*(2), 1-9. https://doi.org/10.1186/1742-4755-12-S2-S5

BBC(2019, March 3). The factors behind the reduction of CS rate in China (Translated from Bangla). https://www.bbc.com/bengali/news-47431700

Becker, M. H., Haefner, D. P., Kasl, S. V., Kirscht, J. P., Maiman, L. A., & Rosenstock, I. M. (1977). Selected psychosocial models and correlates of individual health-related

behaviors. *Medical Care, 15*(5), 27-46. https://doi.org/10.1097/00005650-197705001-00005

Begum, T., Rahman, A., Nababan, H., Hoque, D. M. E., Khan, A. F., Ali, T., & Anwar, I. (2017). Indications and determinants of caesarean section delivery: evidence from a population-based study in Matlab, Bangladesh. *PloS One*, *12*(11), e0188074. https://doi.org/10.1371/journal.pone.0188074

Betran, A. P., Temmerman, M., Kingdon, C., Mohiddin, A., Opiyo, N., Torloni, M. R., ... & Downe, S. (2018). Interventions to reduce unnecessary caesarean sections in healthy women and babies. *The Lancet, 392*(10155), 1358-1368. https://doi.org/10.1016/S0140-6736(18)31927-5

Bollen, K. A., & Long, J. S. (1992). Tests for structural equation models: introduction. Sociological Methods & Research, 21(2), 123-131. https://doi.org/10.1177/0049124192021002001

Buyukbayrak, E. E., Kaymaz, O., Kars, B., Karsidag, A. Y. K., Bektas, E., Unal, O., & Turan, C. (2010). Caesarean delivery or vaginal birth: preference of Turkish pregnant women and influencing factors. *Journal of Obstetrics and Gynaecology, 30*(2), 155-158. https://doi.org/10.3109/01443610903461436

Carpenter, C. J. (2010). A meta-analysis of the effectiveness of health belief model variables in predicting behavior. *Health Communication, 25*, 661–669. https://doi.org/10.1080/10410236.2010.521906

Coates, D., Thirukumar, P., Spear, V., Brown, G., & Henry, A. (2020). What are women's mode of birth preferences and why? A systematic scoping review. *Women and Birth*, *33*(4), 323-333. https://doi.org/10.1016/j.wombi.2019.09.005

Darsareh, F., Aghamolaei, T., Rajaei, M., Madani, A., & Zare, S. (2016). The differences between pregnant women who request elective caesarean and those who plan for vaginal birth based on Health Belief Model. *Women and Birth, 29*(6), e126-e132. https://doi.org/10.1016/j.wombi.2016.05.006

Dey, S. K. (2017, September 18). The cesarean section becomes a fashion for pregnant women (translated from Bangla). https://www.dw.com/bn/ গর্ভবতী-মায়েদের-কাছে-সিজার-এখন-ফ্যাশন/ a-40538725

Doraiswamy, S., Billah, S. M., Karim, F., Siraj, M. S., Buckingham, A., & Kingdon, C. (2021). Physician–patient communication in decision-making about Caesarean sections in eight district hospitals in Bangladesh: a mixed-method study. *Reproductive Health, 18*(1), 1-14. https://doi.org/10.1186/s12978-021-01098-8

Dutta, M. J. (2008). Communicating health: A culture-centered approach. London: Polity Press

Fan, Y., Chen, J., Shirkey, G., John, R., Wu, S. R., Park, H., & Shao, C. (2016). Applications of structural equation modeling (SEM) in ecological studies: an updated review. *Ecological Processes, 5*(1), 1-12. https://doi.org/10.1186/s13717-016-0063-3

Farnworth, A., Robson, S. C., Thomson, R. G., Watson, D. B., & Murtagh, M. J. (2008). Decision support for women choosing mode of delivery after a previous caesarean section: a developmental study. *Patient Education and Counseling*, *71*(1), 116-124.

Glanz, K., Rimer, B. K., & Viswanath, K. (Eds.). (2008). Health behavior and health education: theory, research, and practice. John Wiley & Sons.

Goodall, K. E., McVittie, C., & Magill, M. (2009). Birth choice following primary Caesarean section: mothers' perceptions of the influence of health professionals on decision-

**Cesarean Trends in Bangladesh**                                                                        24making. *Journal of Reproductive and Infant Psychology*, *27*(1), 4-14. https://doi.org/10.1080/02646830801918430

Green, E. C., Murphy, E. M., & Gryboski, K. (2020). The health belief model. The Wiley *Encyclopedia of Health Psychology*, 211-214. https://doi.org/10.1002/9781119057840.ch68

Haider, M. R., Rahman, M. M., Moinuddin, M., Rahman, A. E., Ahmed, S., & Khan, M. M. (2018). Ever-increasing Caesarean section and its economic burden in Bangladesh. *PloS one*, *13*(12), e0208623.

Hair, J. F., Black, W. C., Babin, B. J., & Anderson, R. E. (2010). Multivariate data analysis: A global perspective (7th ed.). Upper Saddle River, NJ: Pearson.

Janz, N. K., & Becker, M. H. (1984). The health belief model: A decade later. *Health Education Quarterly*, 11(1), 1-47. https://doi.org/10.1177/109019818401100101

Karim, F., Ali, N. B., Khan, A. N. S., Hassan, A., Hasan, M. M., Hoque, D. M. E., ... & Chowdhury, M. A. K. (2020). Prevalence and factors associated with caesarean section in four Hard-to-Reach areas of Bangladesh: Findings from a cross-sectional survey. PloS one, 15(6), e0234249.

Kline, R.B., 2015. Principles and Practice of Structural Equation Modeling. The Guilford Press.

Kumar, R., & Lakhtakia, S. (2021). Rising cesarean deliveries in India: medical compulsions or convenience of the affluent?. *Health Care for Women International*, *42*(4-6), 611-635.

Loke, A. Y., Davies, L., & Li, S. F. (2015). Factors influencing the decision that women make on their mode of delivery: The Health Belief Model. *BMC Health Services Research*, *15*(1), 1-12. https://doi.org/10.1186/s12913-015-0931-z

**Cesarean Trends in Bangladesh** 25

Naghibi, S. A., Khazaee-Pool, M., & Moosazadeh, M. (2021). The Iranian version of theory-based intention for cesarean section (IR-TBICS) scale: development and first evaluation. *BMC Pregnancy and Childbirth, 21*(1), 1-11. https://doi.org/10.21203/rs.2.14833/v2

Norman, P., Conner, M., & Bell, R. (1999). The theory of planned behavior and smoking cessation. *Health Psychology, 18*(1), 89. https://doi.org/10.1037//0278-6133.18.1.89

Rai, S. D., Poobalan, A., Jan, R., Bogren, M., Wood, J., Dangal, G., ... & Shahid, F. (2019). Caesarean Section rates in South Asian cities: Can midwifery help stem the rise?. *Journal of Asian Midwives* (JAM).

Rich, A., Brandes, K., Mullan, B., & Hagger, M. S. (2015). Theory of planned behavior and adherence in chronic illness: a meta-analysis. *Journal of Behavioral Medicine, 38*(4), 673-688. https://doi.org/10.1007/s10865-015-9644-3

Rosenstock, I. M. (1974). The health belief model and preventive health behavior. *Health Education Monographs, 2*(4), 354-386. https://doi.org/10.1177/109019817400200405

Rosseel, Y. (2012). lavaan: An R package for structural equation modeling. *Journal of Statistical Software, 48*, 1-36. https://doi.org/10.18637/jss.v048.i02

Sandall, J., Tribe, R. M., Avery, L., Mola, G., Visser, G. H., Homer, C. S., ... & Taylor, P. (2018). Short-term and long-term effects of caesarean section on the health of women and children. *The Lancet*, *392*(10155), 1349-1357. https://doi.org/10.1016/S0140-6736(18)31930-5

Shahabuddin, A., Nöstlinger, C., Delvaux, T., Sarker, M., Delamou, A., Bardají, A., ... & De Brouwere, V. (2017). Exploring maternal health care-seeking behavior of married adolescent girls in Bangladesh: a social-ecological approach. *PloS one, 12*(1), e0169109.

Singh, P., Hashmi, G., & Swain, P. K. (2018). High prevalence of cesarean section births in

**Cesarean Trends in Bangladesh** 26private sector health facilities-analysis of district level household survey-4 (DLHS-4) of India. *BMC Public Health, 18*(1), 1-10.

Shams-Ghahfarokhi, Z., & Khalajabadi-Farahani, F. (2016). Intention for cesarean section versus vaginal delivery among pregnant women in Isfahan: Correlates and determinants. *Journal of Reproduction & Infertility, 17*(4), 230. PMCID: PMC5124342

Suhr, D. (2006). The basics of structural equation modeling. Presented: Irvine, CA, SAS User Group of the Western Region of the United States (WUSS).

Verma, V., Vishwakarma, R. K., Nath, D. C., Khan, H. T., Prakash, R., & Abid, O. (2020). Prevalence and determinants of caesarean section in South and South-East Asian women. Plos one, 15(3), e0229906.

WHO (2021, June 16). Caesarean section rates continue to rise, amid growing inequalities in access: WHO. https://www.who.int/news/item/16-06-2021-caesarean-section-rates-continue-to-rise-amid-growing-inequalities-in-access-who

Wolf, E. J., Harrington, K. M., Clark, S. L., & Miller, M. W. (2013). Sample size requirements for structural equation models: An evaluation of power, bias, and solution propriety. *Educational and Psychological Measurement, 73*(6), 913-934. https://doi.org/10.1177/0013164413495237

Yang, Z. J. (2015). Predicting young adults' intentions to get the H1N1 vaccine: an integrated model. *Journal of Health Communication*, *20*(1), 69-79. https://doi.org/10.1080/10810730.2011.571340